\documentclass[aps,pra,reprint,showpacs,superscriptaddress,floatfix,twocolumn]{revtex4-1}
\usepackage{dcolumn}
\usepackage{amssymb,amsmath}
\usepackage{ifpdf}
\ifpdf
  \usepackage{epstopdf}
\fi
\usepackage{graphicx}

\begin{document}

\raggedbottom
\title{Advanced switching schemes in a Stark decelerator}

\author{Dongdong Zhang}
\email{dongdong.zhang@unibas.ch}
\altaffiliation[\\Present address: ]{Department of Chemistry, University of Basel, Basel, Switzerland}
\affiliation{Fritz-Haber-Institut der Max-Planck-Gesellschaft,
Faradayweg 4-6, 14195 Berlin, Germany} 

\author{Nicolas Vanhaecke}
%\surname{Vanhaecke}
\altaffiliation[Present address: ]{European Patent Office, Patentlaan 2, 2288 EE Rijswijk, The Netherlands}
\affiliation{Fritz-Haber-Institut der Max-Planck-Gesellschaft,
Faradayweg 4-6, 14195 Berlin, Germany}
\affiliation{Laboratoire Aim\'{e} Cotton, CNRS \& Universit\'{e} Paris-Sud, 91405 Orsay, France}

\author{Gerard Meijer}
%\surname{Meijer}
\altaffiliation[Present address: ]{Radboud University Nijmegen, 6265 AJ Nijmegen, The Netherlands}
\affiliation{Fritz-Haber-Institut der Max-Planck-Gesellschaft,
Faradayweg 4-6, 14195 Berlin, Germany}

\date{\today}

\begin{abstract}
We revisit the operation of the Stark decelerator and present a new, optimized operation scheme, which substantially improves the efficiency of the decelerator
at both low and high final velocities, relevant for trapping experiments and collision experiments, respectively. Both experimental and simulation results show that this new mode of operation outperforms the schemes which have hitherto been in use. This new mode of operation could potentially be extended to other deceleration techniques.
\end{abstract}

\pacs{37.10.Mn}

\maketitle
\section{Introduction}

Inspired by their fascinating applications or future applications for precise molecular spectroscopy measurements 
\cite{Amelinkphotoassociation2003,
BasOHlifetime2005,
Jonesphotoassociationsepc2006,
gilijamseCOlifetime2007}, 
quantum information 
\cite{DeMilleQuanComwithTrappedPolarMolelcules2002,
KotochigovaContrPolarMoleinOpticalLattices2006,
ZollerHybridQuanProcessor2006,
ZollerQuantumcomputing2007} 
and reactive or inelastic collisions at low temperatures 
\cite{BasCollisionwithStarkDeceleratedBeam2009,
GianturcoChallengeinColdMol2009,
LudwigStatetoStateinelasticScattering2010,
BasManiandContrMoleBeam2012,
RaizenColdChemistrywithMagDeceleratedMeam2012,
Kirste23112012,
vonzastrow2014,
schewerotationally2015,PhysRevLett.115.133202},
cold molecules have become a very hot topic. 
Many groups all over the world have devoted effort into developing new techniques to generate cold molecules. During the last two decades several methods have been developed, such as Synthesis cooling 
\cite{PilletColdCs2MolePhotoassociation1998,
GozziniColdRuMoleinMOT2000,
EylerKRbPhotoassociation2004,
DeMilleRbCsPhotoassociation2004,
DeMilleRbCsPhotoassociation2005},
 buffer gas cooling 
\cite{WeinsteinBufferGasCooling1999,
DoyleHighFluxBeamSource2005,
DoyleBrightGuidedMoleBeam2007}, 
direct laser cooling \cite{Shuman2010,
PhysRevLett.108.103002,
Barry2014}, velocity selection 
\cite{PhysRevA.67.043406} 
 and Stark or Zeeman deceleration 
\cite{RickStarkDecelerator1999,
RickAGdeceleration2002,
VanhaeckeRydbergDeceleratorusingTravellingElectricTrap2005,
BasTamingMoleBeam2008,
HoganZeemanDecelerator2007,
RaizenAtomicCoilgun2007,
VanhaeckeMagneticTravelingTrap2011}.
In this paper we focus on the Stark deceleration technique. We explore a new operation mode to improve the deceleration efficiency of a Stark decelerator in a large final velocity range.

The technique of Stark deceleration makes it possible not only to arbitrarily vary the final velocities of polar molecules but also to select the molecules in certain quantum states (electronic, vibrational and rotational). That makes this technique realize the full control of molecules. The first realization of the Stark deceleration was in 1999 by Gerard Meijer's group in Nijmegen 
\cite{RickStarkDecelerator1999}.
Previously, a number of molecular species including CO
\cite{RickStarkDecelerator1999}, ND$_{3}$
\cite{RickElectricTrappingNH32000}, OH
\cite{JunPSManipulationofOH2003,
JunColdOHinLabFrame2004,
JunEfficientStarkDecelerationofPolarMole2004,
BasOHlifetime2005,
BasExperimentwithOH2007}, YbF
\cite{TarbuttAGdecelerator2004}, H$_{2}$CO
\cite{JunColdformaldehydeMole2006}, NH
\cite{BasNHDeceleration2006} and SO$_{2}$
\cite{LisdatColdAtomandMolefromFragmentationofDeceleratedSO22006,
GerardColdSO2byStarkDeceleration2008} 
have been successfully decelerated with the Stark deceleration technique. These successes make the Stark decelerator an established method for taming molecular beams and hence enable a greater grange of applications for the beams.
%For example, Stark decelerator can produce completely quantum and movement controlled molecular samples in molecular scattering experiment. Moreover, when the molecules are decelerated to near standstill, they can be loaded into a trap where the observation time can be enhanced to enable the investigation of molecular properties in great details. 

The number of molecules is crucial in experiments using low kinetic energy molecules, such as molecular scattering and trapping. 
High densities of decelerated molecules will improve the precision of metrology experiments and are a prerequisite for future applications of cooling schemes such as evaporative cooling, in order to reach the molecular BEC regime 
\cite{BaranovUltracoldDipolarGases2002}. 
In most Stark deceleration experiments to date, a large drop of molecular density has been found when the final molecular beam velocity is very low 
\cite{JunEfficientStarkDecelerationofPolarMole2004,
BasHigherOrderResonanceinStarkDecelerator2005,
BasTransverseStabilityinStarkDecelerator2006,
JoopOptimizingtheStarkDeceleratorusingEvolutionary2006,
JunMitigationofLossinMolecularStarkDecelerator2008,
LudwigS3OperationMode2009}. 
The loss mechanism at low velocity has already been discussed in detail 
\cite{BasTransverseStabilityinStarkDecelerator2006,
JoopOptimizingtheStarkDeceleratorusingEvolutionary2006,
JunMitigationofLossinMolecularStarkDecelerator2008,
LudwigS3OperationMode2009}. 
In order to overcome the loss mechanism at low final velocity, alternative operation schemes have been proposed, for example: constructing a longer Stark decelerator and operating in the so-called $s$=3 mode or constructing a more complicated decelerator
\cite{LudwigS3OperationMode2009}. 
The longer decelerator operated in the $s$=3 mode demonstrated the most successful scheme so far to reduce the loss at high final velocities(above 100\,m/s), for which every three pairs of electrodes are used together for deceleration, while extra transverse focusing is provided by the intermediate stages. The performance of this scheme beats the conventional $s$=1 operation mode at final velocity lager then 100\,m/s, but at lower final velocities the number of molecules drops even faster than the normal $s$=1 mode
\cite{LudwigS3OperationMode2009}. 

In this paper we revisit the mechanism causing the loss of molecules at very low velocity and propose an optimized high voltage switching sequence to minimize the loss at low final velocity. Both experimental and simulation results show that by applying our switching sequence, we gain a factor of about 2-2.5 in the density of molecules at final velocities as low as 28\,m/s compared to the normal $s$=1 deceleration mode. The longitudinal temperature is also lower compared to the one achieved using a standard deceleration switching  sequence. At high final velocity we obtain intensities and temperatures of the molecular beam comparable to that achieved in an $s$=3 decelerator, but without requiring an increase in decelerator length. This optimal switching sequence scheme is well suited for applications using Stark decelerators. It is easy and cheap to realize and performs better than the conventional way of operating a Stark decelerator. Furthermore, this method can be extended to the operation of Zeeman decelerators which all based on the same principle of functioning. We believe this method will become the new standard routine to operate a decelerator
\cite{katrin2014zeeman,
prakatrin2015,
vogelsoptimal2015}.

\section{Standard switching}\label{sec:standardswitching}

For consistency, let us first recall how longitudinal and transverse properties of the Stark decelerator have been investigated in earlier studies.
We also use this section to introduce the formalism that we will extend to a much larger class of switching sequences in section \ref{sec:advancedswitching}.
The one-dimensional theory of phase-space stability in a Stark decelerator has been described in early works on deceleration of polar molecules \cite{Bethlem:PRA65:053416}, inspired by pioneering works on charged particle accelerators \cite{PhysRev.68.143,JPhysUSSR:Veksler}.
A model of the three-dimensional dynamics in the Stark decelerator has been described in earlier works \cite{BasTransverseStabilityinStarkDecelerator2006}, on which we base the present study.

A Stark decelerator consists of a long array of pairs of electrodes extending in the direction of the molecular beam. 
The electrode pairs are separated by a length $L$ from each other and can be charged to various voltages. 
A periodic electric potential is realized by charging every even pair of electrodes to opposite high voltages and grounding every odd pair of electrodes.
This electric field configuration provides any polar molecule a 2$L$-periodic Stark potential along the molecular beam axis $\hat{z}$, referred to as $W^{(1)}$.
By exchanging the roles of odd and even electrode pairs, one realizes another Stark potential, $W^{(2)}$.
Practically, this is done by switching on and off the necessary voltages.
On the molecular beam axis, $W^{(2)}$ is simply $W^{(1)}$ translated along the molecular beam axis by half a period, \emph{i.e.}, by $L$.
Off-axis, however, the new potential is not only shifted but also rotated by 90$^\circ$ around the molecular beam axis, due to the 90$^\circ$ alternate orientation of the electrode pairs. 

%Small paragraph introducing $\phi_0$ and the field configurations

In the standard mode of operation of a Stark decelerator, for example molecules climb up one Stark potential hill originating from $W^{(1)}$. 
The potential is switched off abruptly, such that the kinetic energy that has been converted into potential energy is not regained. 
At the same time the potential $W^{(2)}$ is switched on, and the process repeats, alternating between $W^{(1)}$ and $W^{(2)}$ until the final velocity is attained. 
The switching sequence applied to the decelerator is calculated with the help of a fictituous particle, called the synchronous particle, which is always at the same position $z_0$ relative to the potential at the moment the fields are switched.
This position is related to the dimensionless phase angle, defined by $\phi_0 = \pi\,z_0 /L$.
By definition, the synchronous particle travels exactly one stage of the decelerator during the time $\Delta T$ between two switching times.
Therefore, the synchronous molecule loses a constant amount of kinetic energy per stage.

\begin{figure}[!htb]
    \centering
    \resizebox{0.9\linewidth}{!}
    {\includegraphics{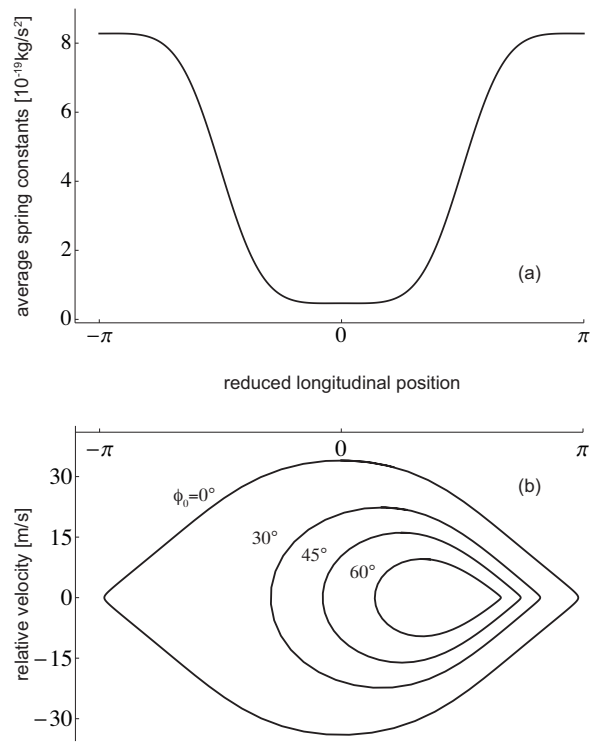}}
    \caption{Transverse and longitudinal properties of the Stark decelerator in its normal mode of operation ($s$=1). The inset (a) shows the average transverse spring constant as a function of the reduced longitudinal position, relative to the position of the $\phi_0=0^\circ$ synchronous particle. The inset (b) shows the separatrix in the longitudinal phase space for various phase angles.
}
    \label{fig:1}
\end{figure}

The Stark energy $W^{(1,2)}$ of a polar molecule (in a low-field-seeking quantum state) on the molecular beam axis in a given field configuration (either 1 or 2) of the Stark decelerator is symmetric around the position of a pair of electrodes. It can be expanded as a Fourier series, with the help of the so-called reduced longitudinal position $\theta = \pi\,z /L \left[ 2\pi \right]$:
\begin{equation}\label{eq:spatial-fourier}
\begin{split}
W^{(1)}(\theta)
& = a_0 + \sum_{n=1}^{\infty} a_n \cos(n( \theta + \pi/2))\\
& = a_0 - a_1 \sin\theta - a_2 \cos 2\theta + a_3 \sin 3 \theta  + ...
\quad ,
\end{split}
\end{equation}
and 
\begin{equation}
W^{(2)}(\theta)=W^{(1)}(\theta-\pi) \quad .
\end{equation}
As long as the change in Stark energy of a molecule is small compared to the kinetic energy of the molecule, the dynamics of the molecules throughout the decelerator are slow on the time scale of $\Delta T$. They are well described by a time-averaged force over the different field configurations, taken over two stages of the decelerator.
Due to the negligible change in velocity over two stages, the time average can be replaced by the spatial average over the (reduced) longitudinal position.
For the synchronous molecule, the averaged longitudinal force reads:
\begin{eqnarray}\label{eq:fbarnormal1}
\nonumber
\bar{F}_z ^{\phi_0}
&=&
\frac{1}{2\pi}
\left[
\int_{\phi_0}^{\phi_{1}}
F_z^{(1)}(\theta)
d\theta
+
\int_{\phi_1}^{\phi_{2}}
F_z^{(2)}(\theta)
d\theta \,,
\right]
\\ \nonumber
&=& \frac{1}{2 L}
\left[
-W^{(1)}(\phi_1)
+W^{(1)}(\phi_0) \right.
\\ \nonumber
&&\left. \hspace{3cm}
-W^{(2)}(\phi_2)
+W^{(2)}(\phi_1)
\right] 
\\ 
&=& 
-\frac{2a_1}{L} \sin \phi_0 \,,
\end{eqnarray}
using $\phi_1=\phi_0+\pi$ and $\phi_2=\phi_0+2\pi$, and the standard definition of the phase angle $\phi_0 = \pi z_0 /L $.
%The synchronous molecule is the (fictitious) molecule that is always in phase with the fields that are being switched, \emph{i.e.}, its reduced position is always $\phi_0$ after the fields have been switched.
For a non-synchronous particle, which has a reduced longitudinal position $\Delta \phi$ relative to that of the synchronous particle, the average force reads

\begin{eqnarray}\label{eq:fbarnormal2}
\\ \nonumber
\bar{F}_z ^{\phi_0}(\Delta\phi)
&=&
\frac{1}{2\pi}
\left[
\int_{\phi_0+\Delta\phi}^{\phi_{1}+\Delta\phi}
F_z^{(1)}(\theta)
d\theta
\right.
\\ \nonumber
&&\left. \hspace{2cm}
+
\int_{\phi_1+\Delta\phi}^{\phi_{2}+\Delta\phi}
F_z^{(2)}(\theta)
d\theta \,,
\right]
\\ \nonumber
&=& 
-\frac{2a_1}{L} \sin \left( \phi_0 +\Delta\phi \right) \, .
\end{eqnarray}

Note that in the frame of this model, any non-synchronous particle travels exactly 2$L$ during 2$T$, just like the synchronous particle.
A higher order approximation has been carried out in \cite{BasHighorderStarkPRA2005} but is not necessary for our present goals.
The phase angle $\phi_0$ at which the decelerator is operated determines both the deceleration rate and the longitudinal acceptance of the decelerator. For a given value of $\phi_0$, molecules that have a position in phase space that is within the acceptance of the decelerator, bound by the so-called separatrix, are phase stable and are selected by the decelerator. This has been described in very early works on the Stark decelerator.
Figure \ref{fig:1}(b) shows the separatrix in the longitudinal phase space for several phase angles $\phi_0$. All separatrices are given for OH radicals in the upper $\Lambda$-doublet component of the rovibronic ground state. 

Usually, the dynamics in the longitudinal phase space are described using only one longitudinal potential, and by averaging over only one stage of the decelerator.
Both field configurations are nevertheless necessary to describe the transverse properties of the Stark decelerator, and the average properties can only be calculated by averaging over at least one full geometrical period of the decelerator, \emph{i.e.}, two stages. 
In our model, several approximations on the transverse dynamics are made.  
First, the transverse dynamics is slow on the time scale needed to travel over 2$L$ at a velocity of $v_{\rm sync}$. Therefore any change in transverse directions after travelling 2$L$ is neglected in the calculation of the transverse forces.
Second, we consider the transverse properties close to the molecular beam axis, where the transverse force is to first order approximated by a linear restoring force:
\begin{equation}
F_{x}^{(i)}(x,y,\theta) = k_{x}^{(i)}(\theta) x \quad ,
\end{equation}
where $i$ denotes again the field configuration, and $\theta$ the reduced longitudinal position. A similar expression holds for the $y$-component.
The average transverse forces are therefore also linear restoring forces with restoring spring constants $\bar{k}_{x,y}$.
These constants can be written in a manner similar to that used for the average longitudinal force:
%Therefore the average transverse restoring constants can be written in a similar way as the average longitudinal force:
%The average transverse restoring constants are calculated following the same procedure as for the longitudinal force: 
\begin{eqnarray}
\\ \nonumber
\bar{k}_x ^{\phi_0}(\Delta\phi)
&=&
\frac{1}{2\pi}
\left[
\int_{\phi_0+\Delta\phi}^{\phi_{1}+\Delta\phi}
k_x^{(1)}(\theta)
d\theta
\right.
\\ \nonumber
&&\left. \hspace{2cm}
+
\int_{\phi_1+\Delta\phi}^{\phi_{2}+\Delta\phi}
k_x^{(2)}(\theta)
d\theta 
\right] \quad ,
\end{eqnarray}
and the same expression holds for the $y$-component.
The spring constants can be evaluated numerically, and depend on the phase angle $\phi_0$. 
They have been expressed equivalently in terms of average natural frequencies in previous studies \cite{BasTransverseStabilityinStarkDecelerator2006}. 
Figure \ref{fig:1}(a) depicts the average transverse spring constant as a function of the reduced longitudinal position of a molecule (\emph{i.e.} $\phi_0+\Delta\phi$).
As can be seen, this dependence is rather pronounced, and has been shown to be responsible for couplings between longitudinal and transverse motions, and poor focusing properties at low phase angles. Both effects induce unwanted losses throughout the deceleration process \cite{BasTransverseStabilityinStarkDecelerator2006}. 

To circumvent the unwanted transverse properties of the normal mode of operation, another mode of operation, the so-called $s$=3 mode, has been introduced \cite{BasTransverseStabilityinStarkDecelerator2006}, in which the decelerator is switched only once for each time the synchronous molecule has travels a distance of three stages. This increases the overall acceptance noticeably, by reducing the couplings between the transverse and longitudinal motions and by achieving better focusing properties.

\begin{figure}[!htb]
    \centering
    \resizebox{1.0\linewidth}{!}
    {\includegraphics{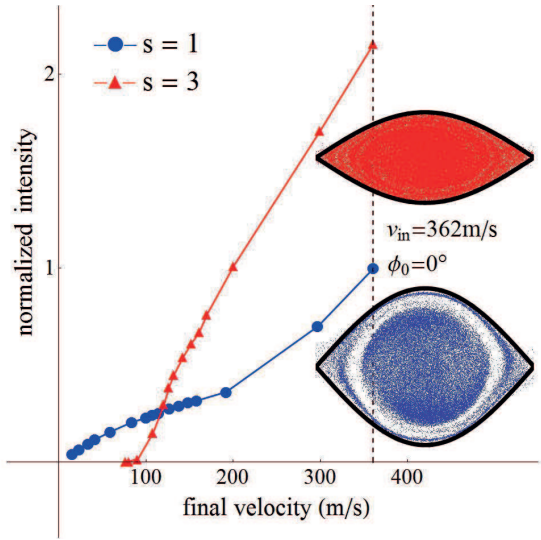}}
    \caption{ Simulated intensity of the decelerated peak as a function of the final velocity, for both $s$=1 (blue dots) and $s$=3 (red triangles) modes of operation of the Stark decelerator. For all data, the initial velocity is chosen to be 362\,m/s. Below 90\,m/s the efficiency of the $s$=3 mode of operation drops to zero. The insets show the longitudinal phase space acceptance of both mode of operations for a phase angle of 0$^\circ$, \emph{i.e.}, a final velocity of 362\,m/s. While the accepted volume in the longitudinal phase space is fully filled within the separatrix with the $s$=3 mode of operation, it exhibits a complex structure with the $s$=1 mode of operation.     
}
    \label{fig:2}
\end{figure}

Let us briefly compare quantitatively the normal mode of operation (called $s$=1) and the $s$=3 modes of operation of a Stark decelerator. 
Figure \ref{fig:2} depicts the intensity of the decelerated peak throughout the decelerator as a function of the final mean velocity of this peak. In all cases the initial velocity is 362\,m/s, and the intensity is normalized to that of the peak guided with $\phi_0$=0$^\circ$ using the $s$=1 mode of operation. 
In addition, the insets of Figure \ref{fig:2} show the longitudinal phase space acceptance for $\phi_0$=0$^\circ$, in both $s$=1 and $s$=3 modes of operation, as well as the separatrices. As shown in previous studies, in the $s$=1 case, the acceptance exhibits a complex structure in the longitudinal phase space, while in the $s$=3 case, the phase space seems to be fully filled inside the separatrix. 
The efficiency of both modes of operation drops when the final velocity is decreased, \emph{i.e.}, with increasing phase angles.
At small phase angles, the $s$=3 mode of operation outperforms the $s$=1 one, thanks to better focusing properties, and because it avoids couplings between longitudinal and transverse motions. 
At larger phase angles, however, lower velocities are reached, and the efficiency of the $s$=3 mode drops to zero below 100\,m/s, due to overfocusing properties of this scheme.

\section{Advanced switching}\label{sec:advancedswitching}
Let us now extend the model presented above to a large, new class of switching schemes. 
We now allow the decelerator to be switched 2$N$ times while the synchronous particle flies a distance corresponding to 2$K$ stages along the beam axis, $K$ and $N$ being integer numbers. We define a set of reduced positions $\{\phi_i\}_{i=0..2N-1}$, which describe the position of the synchronous particle at each of the 2$N$ switching times.
The average longitudinal force reads now, by extension of Equation \ref{eq:fbarnormal2}:
\begin{eqnarray}
\\ \nonumber
\bar{F}_z^{K,N,\{\phi_i\}}
(\Delta\phi)
&=& 
\frac{1}{2\pi K}
\sum_{i=0}^{2N-1}
\int_{\phi_i+\Delta\phi}^{\phi_{i+1}+\Delta\phi}
F_z^{(i[2]+1)}(\theta)
d\theta \,,
\end{eqnarray}
where $i[2]$ stands for $i$ modulo 2. By definition, the series $\{ \phi_i\}_i$ is ordered, \emph{i.e.}, $\phi_i < \phi_{i+1}$, and by definition $\phi_{2N} = \phi_0 + 2\pi K$. For consistency, the reduced position is now defined by $\theta = \pi\,z /L \left[ 2\pi K \right]$.
In this scheme, the synchronous particle fulfills $\Delta\phi\equiv$0, just like in the standard mode of operation.

The average transverse force constants can be expressed following the same formalism:
\begin{eqnarray}
\\ \nonumber
\bar{k}_{x,y} ^{K,N,\{\phi_i\}}
(\Delta\phi)
&=& 
\frac{1}{2\pi K}
\sum_{i=0}^{2N-1}
\int_{\phi_i+\Delta\phi}^{\phi_{i+1}+\Delta\phi}
k_{x,y}^{(i [2]+1)}(\theta)
d\theta \,.
\end{eqnarray}
The set of phase angles $\{\phi_i\}_i$ realises a way of sampling both field configurations available in the decelerator alternatively, in a manner quite similar to the pulse width modulation technique commonly used in electronics \cite{pressman1977switching}.

%Description of fig 2 + pb of s=1 and s=3

\begin{figure}[!htb]
    \centering
    \resizebox{1.0\linewidth}{!}
    {\includegraphics{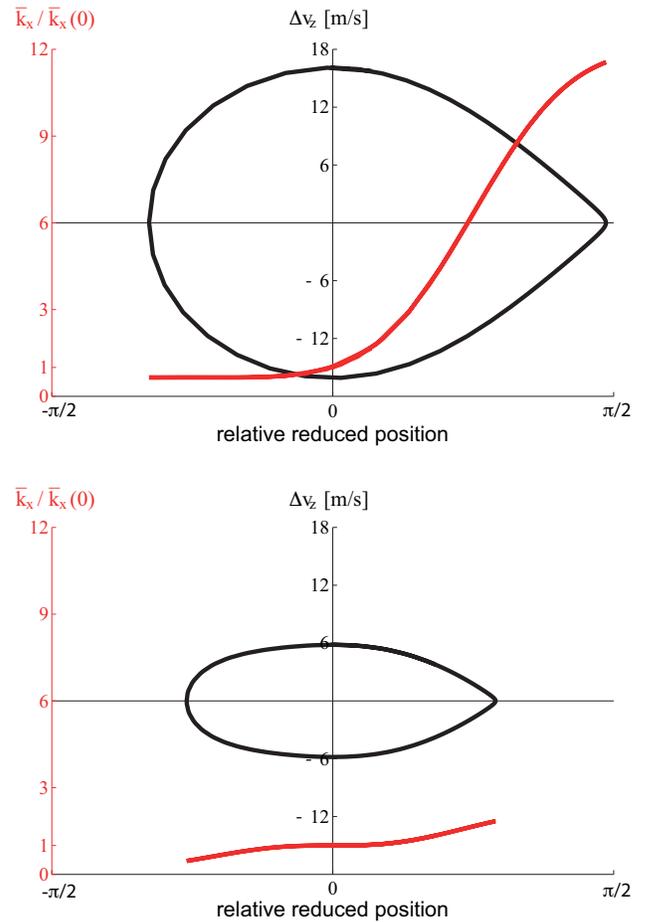}}
    \caption{Longitudinal and transverse properties of the Stark decelerator operated with (a) a standard sequence, and (b) an advanced switching sequence. 
The black solid line depicts the separatrix in the longitudinal phase space, as a function of reduced position and velocity relative to those of the synchronous molecule. The red dashed curve shows the transverse average spring constant, $\bar{k}_x$, as a function of the relative reduced longitudinal position. The transverse average spring constant is normalized to that experienced by the synchronous particle $\bar{k}_x(0)$.
}
    \label{fig:3}
\end{figure}

The standard way of operation of the Stark decelerator corresponds to $(K,N)=(1,1)$, with the additional constraint $\phi_1=\phi_0+\pi$, such that both transverse directions play the same role, and in the end, such that the decelerated packet of molecules have identical properties in both transverse directions.
In that case, denoted $(1,1,\{\phi_0,\phi_0+\pi\})$, the phase angle $\phi_0$ is the only control knob, which therefore determines the rate of deceleration, the acceptance in the longitudinal phase space, as well as the transverse properties, \emph{i.e.}, the acceptance in the transverse phase spaces.
 The $s$=3 mode of operation is described in the present formalism by $(K,N)=(3,1)$, with an additional constraint $\phi_1=\phi_0+3\pi$. Here again, the choice of $\phi_0$ determines all longitudinal and transverse properties of the decelerated packet of molecules.

On the contrary, if $N \geq 2$, there is generally not a \emph{unique} set of phase angles $\{\phi_i\}_i$ that allows the removal of a given amount of kinetic energy over 2$K$ stages, but many such sets of phase angles. 
This opens the possibility to search for sequences with additional tunable properties, for example in terms of longitudinal and transverse acceptance.
Alternatively, or in addition, one could search for sequences that provide average spring constants $\bar{k}_{x,y}(\Delta\phi)$ closely matching a given function of reduced longitudinal position $\Delta\phi$.

Figure \ref{fig:3} shows the comparison of longitudinal and transverse properties of a standard mode of operation (a) with an advanced switching scheme (b).
In both cases, the separatrix in the longitudinal phase space is represented by a solid black line, as a function of the relative position in phase space, for which the origin has been chosen to be the synchronous molecule. 
Dashed red curves depict the average transverse spring constants $\bar{k}_{x,y}(\Delta\phi)$ as a function of the relative reduced longitudinal position $\Delta\phi$, normalized to the average transverse spring constant experienced by the synchronous molecule $\bar{k}_{x,y}(0)$.

In case (a), a standard switching is used with a phase angle of $45^\circ$, which determines all properties of the switching scheme. 
Already visible in Fig.~\ref{fig:1}, the fact that the transverse spring constant exhibits a strong dependence on the reduced longitudinal position is emphasized in Fig.~\ref{fig:3}.
The spring constant is more than one order of magnitude stronger for molecules far ahead of the synchronous molecules than for molecules lying behind. 
This is a major loss mechanism during the deceleration process: as a given molecule revolves around the synchronous molecule in the longitudinal phase space, it gets correctly focused when it flies ahead of the synchronous molecule, but badly when it flies behind it. Eventually the molecule becomes lost.  
Our case studies of simulated trajectories have confirmed this fact, and shown that it is especially pronounced at low velocities. %why low?  
The transverse acceptance is roughly defined by the lowest values of the transverse spring constant.

In case (b), we use an advanced scheme with $(K,N)=(1,3)$. The set of six phase angles is optimized with respect to several goals. 
First, the amount of kinetic energy taken from the synchronous particle is required to be equal to that removed in the standard switching mode of case (a): $(K,N,\{\phi_0,\phi_1\})$=$(1,1,\{45^\circ,225^\circ\})$. Conditions are therefore required on $\bar{F}_{z} ^{K,N,\{\phi_i\}}(\Delta\phi\!=\!0)$.
Second, we require the scheme to ensure longitudinal phase space stability. This is done by requiring the first derivative of the average longitudinal force $\frac{d}{d\Delta\phi} \bar{F}_{z} ^{K,N,\{\phi_i\}} (\Delta\phi\!=\!0)$ to be negative, such that particles ahead of the synchronous one are more decelerated than the synchronous particle, while particles behind it are less decelerated. 
Third, in order to avoid couplings between longitudinal and transverse couplings, both average transverse spring constants $\bar{k}_{x}$ and $\bar{k}_{y}$ are required to be as independent as possible of the reduced longitudinal position. To do so, conditions are set on the first derivative of the average transverse spring constants $\frac{d}{d\Delta\phi} \bar{k}_{x,y} ^{K,N,\{\phi_i\}} (\Delta\phi\!=\!0)$.
The set of phase angles is optimized in a least square fitting procedure with a fitness function which includes our different goals with various weighting factors. 

The optimized set of phase angles $\{\phi_i\}_{i=0..5}$ for case (b) reads 
$\{
 56.4^\circ,
 91.8^\circ,
115.0^\circ,
236.4^\circ,
271.8^\circ,
295.0^\circ\}$. 
%Back to fig3
As can be seen in Figure \ref{fig:3}, the longitudinal acceptance is strongly reduced by using the advanced scheme compared to the normal scheme. However, the transverse force is clearly less strongly dependent on the longitudinal reduced position, which was one of our goals. 
While the average transverse spring constant differs by a factor 18 within the separatrix in the standard case (a), it does not even differ by a factor 3 within the separatrix in the advanced switching scheme (b). 
%While the average transverse spring constant spans an interval of 1 to 18 in the standard case (a), it spans only a   
In case (b) the optimization of the six phase angles is ran with identical goals for both transverse directions. This symmetry is reflected by the optimized switching scheme which exhibits a systematic relationship between the phase angles: $\phi_{i+3}=\phi_{i}+\pi$.

% comparison $\bar{k}_{x,y}(0)$ in cases a and b

\section{Experiment}
A detailed description of the machine we used in our experiment has been given previously 
\cite{BasOHlifetime2005,
BasHigherOrderResonanceinStarkDecelerator2005,
BasTransverseStabilityinStarkDecelerator2006,
BasNHDeceleration2006,
BasExperimentwithOH2007}. 
We choose OH radicals as the benchmark molecule to investigate the performance of the new class of switching sequences. 
A pulsed beam of OH radicals is generated by photodissociation of HNO$_{3}$ co-expanded with Xe through a pulsed solenoid valve (General Valve Series 99). 
The dissociation takes place inside a quartz capillary that is mounted on the nozzle orifice. 
The 193\,nm laser beam is focused onto the tip of the quartz capillary to make sure that the OH packet is generated in a well-defined timing and position. 
The mean velocity of the molecular beam is around 362\,m/s with a longitudinal velocity spread of 15\% (full width at half maximum). 

After the supersonic expansion, most of the OH radicals in the beam are in the lowest rotational and vibrational state of the $X^{2}{\Pi}_{3/2}$ electronic ground state. 
This $J$=3/2 level has an ${\Lambda}$-doublet splitting of 1.6GHz. 
In the presence of an applied electric field, the upper $\Lambda$-doublet component (of spectroscopic parity $f$) splits into a $M_{J}{\Omega}$=-3/4 and a $M_{J}{\Omega}$=-9/4 states, which are both low field seeking states. 
%Since the Stark energy of $M_{J}{\Omega}$=-9/4 is larger than the other component, OH radicals only in this state will be take into account and decelerated. 
The OH beam passes through a 2-mm-diameter skimmer and enters a second vacuum chamber where the OH molecules are focused into the Stark decelerator by a 37-mm-long hexapole guide. 
The Stark decelerator consists of 109 pairs of electrodes with every pair perpendicular to the previous one, and a center to center distance of 11mm. 
Each electrode is a 6-mm-diameter cylindrical rod, and the center to center distance of the two electrodes of each pair is 10\,mm. This provides a 4${\times}$4mm$^{2}$ aperture to the molecular beam. 
All electrodes are arranged with an angle of 45$^{\circ}$ with respect to the horizontal plane of the laboratory.
% to minimize the bias introduced by the gravity. 
The decelerator is operated with a voltage difference of 40\,kV between opposite electrodes of each pair, which creates a maximum electric field strength on the molecular beam axis of around 91\,kV/cm. 
%The individual electrodes of the decelerator are electrically connected to a high voltage switch (Behlke Lelktronik HTS 301-03-GSM). 
%A total number of four independent switches are requested for the decelerator. 
A set of four independent high voltage switches (Behlke Elktronik HTS 301-03-GSM) allows us to drive the required voltages on all electrodes and therefore to alternatively generate both usual electric field configurations used in a Stark decelerator \cite{Bethlem:PRA65:053416} as discussed in the previous section.

The decelerated molecular beam crosses at right angle with a 282\,nm laser beam about 21\,mm downstream from the end of the decelerator. 
The OH radicals are excited using the $Q_{1}(1)$ transition of the $A ^{2}{\Sigma}^{+}$, $v$'=1 $\leftarrow$ $X ^{2}{\Pi}_{3/2}$, $v$"=0 band. The resulting off-resonant fluorescence of the $A ^{2}{\Sigma}^{+}$, $v$'=1 $\rightarrow$ $X ^{2}{\Pi}_{3/2}$, $v$"=1 band around 313\,nm is collected by a photomultiplier tube. 
In order to minimize the signal fluctuations induced by the shot-to-shot laser intensity fluctuations, we use a laser intensity which always high enough to saturate the $ Q_{1}(1)$ transition around 282\,nm.
Stray light coming from the laser pulse is reduced by passing the laser beam through carefully aligned light baffles and by introducing optical filters in front of the photomultiplier tube. 

In the experiments presented in this article using an advanced switching scheme with $N$$>$$K$, one requires the switches to operate more often than in standard deceleration sequences.
Care has to be taken of the amount of heat that is produced in the switches, which can be an experimental limitation in the applications of advanced switching sequences. 
For the results presented here, we mostly use a sequence with $(K,N)=(1,3)$, which requires three times more switching than a normal sequence. 
The temperature of the switches is monitored during operation, and in order to stay on the safe side, we use a reduced repetition rate of 5\,Hz (instead of 10\,Hz).
Nevertheless, a oil-cooling system implemented on the switches would allow running the experiment at 10\,Hz without generating too much heat.

\begin{figure}[!htb]
    \centering
    \resizebox{\linewidth}{!}
    {\includegraphics{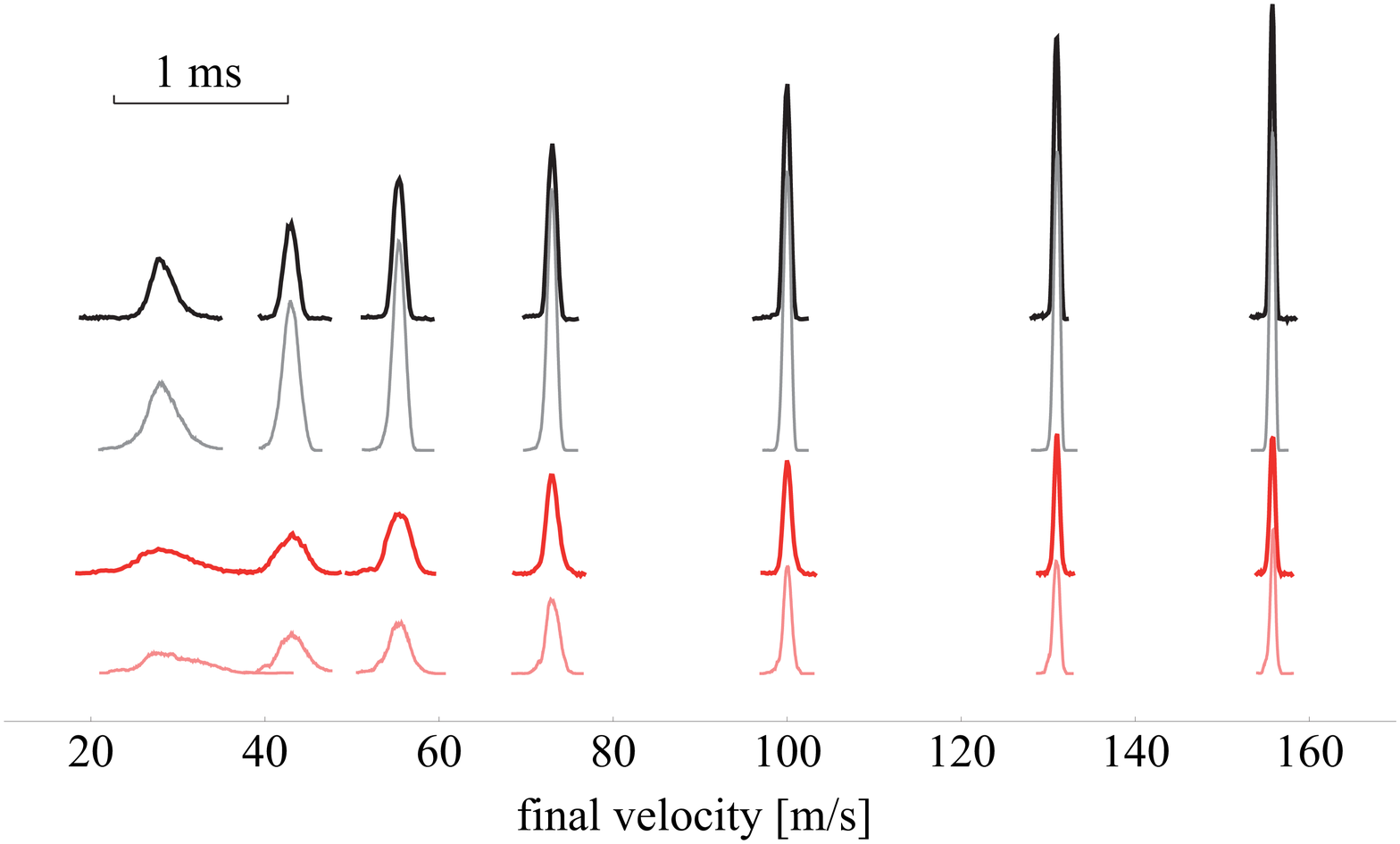}}
    \caption{ Excerpts of time-of-flight profiles obtained with standard sequences (lowest two traces) and advanced sequences (upper traces). Only the decelerated peaks are shown, and are centered on the final velocity of the synchronous particle, which takes the following values: 
28.1\,m/s,
43.0\,m/s,
55.4\,m/s, 
73.0\,m/s, 
100.0\,m/s, 
131.0\,m/s, and
155.8\,m/s
.
In the top-left corner the horizontal time scale is shown, which is common to all time-of-flight profiles. For both types of sequences the experimental data (upper trace) are compared to the output of trajectory simulations (lower trace).}
% The final velocities of the synchronous particle read: 155.8, 131.0, 100.0, 73.0, 55.4, 43.0, 28.1\,m/s}
    \label{fig:4}
\end{figure}

\begin{table}
\begin{tabular}{r@{.}c|cccccc}
\hline
\multicolumn{2}{c|}{$v_{\textrm{final}}$}& \multicolumn{6}{c}{ $\{\phi_i\}_{i=0..5}$ }\\
\hline
155&8\,m/s& 51.2$^\circ$& 92.3$^\circ$& 117.7$^\circ$& 231.2$^\circ$& 272.3$^\circ$& 297.7$^\circ$\\
131&0\,m/s& 53.3$^\circ$& 92.2$^\circ$& 116.9$^\circ$& 233.3$^\circ$& 272.2$^\circ$& 296.9$^\circ$\\
100&0\,m/s& 55.2$^\circ$& 92.0$^\circ$& 115.8$^\circ$& 235.2$^\circ$& 272.0$^\circ$& 295.8$^\circ$\\
73&0\,m/s&  56.4$^\circ$& 91.8$^\circ$& 115.0$^\circ$& 236.4$^\circ$& 271.8$^\circ$& 295.0$^\circ$\\
55&4\,m/s&  57.0$^\circ$& 91.8$^\circ$& 114.6$^\circ$& 237.0$^\circ$& 271.8$^\circ$& 294.6$^\circ$\\
43&0\,m/s&  57.3$^\circ$& 91.8$^\circ$& 114.4$^\circ$& 237.3$^\circ$& 271.8$^\circ$& 294.4$^\circ$\\ 
28&1\,m/s&  57.6$^\circ$& 91.7$^\circ$& 114.2$^\circ$& 237.6$^\circ$& 271.7$^\circ$& 294.2$^\circ$\\
\end{tabular}
\caption{Details of the different sets of phase angles used in the experiment and in the trajectory simulations presented in Fig.\ref{fig:4}.}
\end{table}

\section{Results and discussion}
%% Fig 5
Figure \ref{fig:4} shows excerpts of measured and simulated time-of-flight profiles, showing the decelerated part of the beam only, both for normal sequences (lowest two traces) and advanced sequences (upper traces).
Dark lines are measured time-of-flight profiles, and lighter traces are the output of three-dimensional trajectory simulations.
The different time-of-flight profiles are horizontally centered on the final velocity of the synchronous molecule, and use the same horizontal time scale (shown horizontally in the top left corner) but different time origins. 
Note that for identical sets of initial and final velocities of the synchronous molecule, normal and advanced sequences give rise to decelerated peaks that arrive at noticeably different times. Especially visible at low final velocities, this effect is due to the details of the sequences, which are different enough to induce visibly different arrival times of the decelerated packets. 

Experimental time-of-flight profiles agree very well with the results of our simulations, both for normal and advanced sequences. The highest discrepancy between experiment and simulation occurs for the advanced sequence at low phase angles, \emph{i.e.}, at high final velocities. This discrepancy is attributed to the role of rise and fall times of the electric fields. Indeed, when molecules travel quickly in a decelerator employing an advanced sequence (\emph{e.g.} switching three times per stage), they experience electric field which are rising or falling for a relatively large amount of their flight time. 
%This effect is reduced when the final velocity is lower, because the relative amount of flying time spent feeling rising or falling fields scales linearly with the velocity. 
The improvement of the advanced sequence over the normal sequence in terms of the TOF signal is clearly visible in Figure~\ref{fig:4}. This can be quantized by integrating the signal over the decelerated peak, with an appropriate correction to account for the fact that molecules flying slower spend more time in the detection volume and therefore contribute more to the signal. We find that the gain in signal is about a factor of 2$-$2.5 over the whole range of final velocities explored in Figure~\ref{fig:4}. 

The gain of the advanced sequence over the normal sequence is especially striking at very low final velocities. Indeed, as was already expected from Figure~\ref{fig:3}, beams decelerated with the advanced sequence are much colder in the longitudinal direction than beams decelerated with the standard sequence. At low final velocities, the decelerated packets have enough time to spread between the end of the decelerator and the detection zone, and the ballistic expansion becomes visible on the shape of the detected peak. Colder beams expand less and give rise to a more pronounced peak, as it is clearly visible for a final velocity of 28.1\,m/s using the advanced switching sequence.

The properties of decelerated beams can be compared more quantitatively by analysing the results of the three-dimensional simulations. 
The deceleration of OH radicals from 362\,m/s down to 28.1\,m/s using a normal sequence leads to a decelerated peak with a longitudinal temperature of 56\,mK and transverse temperatures of 12\,mK. With the advanced sequence used in Figure~\ref{fig:4}, the longitudinal temperature drops to 24\,mK and the transverse temperature rise to 22\,mK. Overall, the beam is colder and has much more homogeneous temperatures, which could match better the phase space acceptance of a trap.

Three-dimensional simulations also offer the opportunity to evaluate the phase space acceptance of the different sequences, by calculating the six-dimensional phase space volume of the decelerated packets.
To do so, we use a multi-dimensional numerical integration routine using Monte-Carlo methods (C\textsc{\footnotesize UBA} library) \cite{Hahn2006273}.
This is almost compulsory when the multi-dimensional volume exhibits complicated structures, as is indeed the case for packets decelerated with the standard sequence.
We find that the normal sequence used to decelerate from 362~m/s down to 28.1~m/s has a phase space acceptance of 23$\times 10^3$~mm$^3$(m/s)$^3$. The advanced sequence used in Figure~\ref{fig:4} to achieve the same final velocity of 28.1~m/s has a more than twice as large phase space acceptance, with a volume of 50$\times 10^3$~mm$^3$(m/s)$^3$.

The type of switching sequences we have presented above opens new avenues in the manipulation of molecular beams in a decelerator, and can find applications in various types of experiments.
In state-of-the-art collision studies with crossed molecular beams, it has become crucial to achieve a high collision energy resolution. This is a prerequisite for observing resonances and quantum effects in cold collisions, and the Stark decelerator is one of the few tools to achieve this \cite{Gilijamse:Science313:1617, Kirste23112012}. 
An advanced sequence of the type described above could be optimized to produce a beam with low temperatures in both directions of the crossed-beams plane, \emph{i.e.} longitudinally and in one transverse direction, but rather hot with a large acceptance in the transverse direction that is perpendicular to the collision plane.
Compared to the use of a standard sequence, maybe followed by a spatial filter, such an advanced sequence would enhance the collision energy resolution while maintaining the highest possible flux of molecules out of the decelerator.

Another desired property of state-of-the-art molecular beams is state-purity. This is of great importance in (state-to-state) collision studies, but also for molecular beams that are slowed down to be loaded into a trap. The advanced switching scheme could also be optimized to decelerate and capture a given quantum state, and at the same time avoid to capture (an)other, undesired quantum state(s).

The advanced switching sequences could also be of great impact on the efficiency of trap loading. In all trapping experiments using decelerated molecular beams, the decelerators are operated in a standard mode till the last stage of the decelerator. 
To improve the trap loading efficiency, several strategies have been followed, such as optimization of the number of trapped molecules with self-learning algorithms \cite{JoopOptimizingtheStarkDeceleratorusingEvolutionary2006} and specific design of the trap electrodes \cite{joop2010}. 
Nevertheless, as we have pointed out above, the standard mode of operation has no additional degree of freedom than its (single) phase angle, and there is no way of tuning the longitudinal and transverse properties of the beam exiting the decelerator once the final velocity has been fixed. 
On the contrary, switching the field configurations multiple times in the last few stages of the decelerator would allow one to tune the phase space properties of the beam, prior to loading them into the trap, while nonetheless achieving the required final velocity at the exit of the decelerator.
Since the molecules spend a lot of time in the last few stages of the decelerator, one could afford to switch many times per stage without any experimental limitation. 
Note that the concept of averaged force breaks down at these very low velocities, and one should therefore examine and optimize the dynamics of the molecules in the last few stages of the decelerator from three-dimensional trajectory simulations.

The advanced switching sequences in the Stark decelerator are not limited to toggling between two field configurations. For example, one could easily use a third configuration, where all fields are off. This opens even more possibilities for tuning the phase-space properties at will. Nevertheless, care should be taken of the fact that non-adiabatic transitions might occur when the fields are completely switched off \cite{Kirste-PRA79-051401}. A straightforward way to circumvent this issue is to implement a small bias field, which does not give rise to any substantial force, but keeps the quantum states of interest non-degenerate.

Finally, the strategy of optimizing multiple switching times over several stages can be extended to the Zeeman deceleration technique. Note that the experimental limitation might be more severe in a Zeeman decelerator than in a Stark decelerator, due to the longer rise and fall time experimentally achievable \cite{HoganZeemanDecelerator2007}.

\begin{acknowledgments}
We gratefully acknowledge the help of the electronic laboratory of the Fritz Haber Institute. This work has been funded by the ERC-2009-AdG under Grant Agreement No. 247142-MolChip. D.Z. thanks prof. Gerard Meijer for his supports.
\end{acknowledgments}

\bibliographystyle{apsrev4-1}
\bibliography{bib}

\end{document}